\documentclass[12pt]{article}
\usepackage{amssymb, euscript}
\usepackage{amsmath, bbm}

\usepackage{theorem}
\theoremheaderfont{\scshape}
\theoremstyle{plain}

\newenvironment{definition}[1][Definition]{\begin{trivlist}
\item[\hskip \labelsep {\bfseries #1}]}{\end{trivlist}}

\newcommand{\haken}{\mathbin{\hbox to 8pt{%
                 \vrule height0.4pt width7pt depth0pt
                 \kern-.4pt
                 \vrule height4pt width0.4pt depth0pt\hss}}}

\newcommand{\be}[3]{\begin{equation}  \label{#1#2#3}}
\newcommand{\bea}[3]{\begin{eqnarray}  \label{#1#2#3}}
\newcommand{\ee}{\end{equation}}
\newcommand{\ba}{\begin{array}}
\newcommand{\ea}{\end{array}}
\newcommand{\eea}{\end{eqnarray}}

\newcommand{\Ann}{\mathrm{Ann}\,}
\newcommand{\R}{ {\cal R} }

\begin{document}

\baselineskip=20pt
\parskip=6pt

%%%%%%%%%%%%%%%%%%%%%%%%%%%%%%%%%%%%%%%%%%%%%%

\thispagestyle{empty}

\begin{flushright}
%%%% insert the HU-EP and hep-th number. For the last submit
%%%% the paper, take the # and then resubmit
\hfill{HU-EP-04/71} \\
\hfill{hep-th/0412181}
%\hfill{HU-EP-04/44} \\
%\hfill{hep-th/0408169}%\\
%\hfill{version 08/21/20:45}

\end{flushright}

\vspace{10pt}

\begin{center}{ \LARGE{\bf
Isotropic A-branes and \\[4mm] the stability condition
}}

\vspace{35pt}

{\bf Stefano Chiantese}

\vspace{15pt}

{\it  Humboldt-Universit\"at zu Berlin,\\
Institut f\"ur Physik,\\
Newtonstra\ss e 15, 12489 Berlin, Germany,\\
E-mail: chiantes@physik.hu-berlin.de.}\\[1mm]

\vspace{8pt}

\vspace{40pt}

{\bf ABSTRACT}

\end{center}

The existence of a new kind of branes for the open topological
A-model is argued by using the generalized complex geometry of Hitchin
and the SYZ picture of mirror symmetry. Mirror symmetry suggests to
consider a bi-vector in the normal direction of the brane and a
new definition of generalized complex submanifold. Using this
definition, it is shown that there exists generalized complex
submanifolds which are isotropic in a symplectic manifold. For
certain target space manifolds this leads to isotropic A-branes,
which should be considered in addition to Lagrangian and
coisotropic A-branes. The Fukaya category should be enlarged with
such branes, which might have interesting consequences for the
homological mirror symmetry of Kontsevich. The stability condition
for isotropic A-branes is studied using the worldsheet approach.

 \noindent

\vfill

\newpage

%%%%%%%%%%%%%%%%%%%%%%%%%%%%%%%%%%%%%%%%%%%%%%%%%%%%%%

\section{Introduction}

%%%%%%%%%%%%%%%%%%%%%%%%%%%%%%%%%%%%%%%%%%%%%%%%%%

Since the seminal paper of Witten \cite{Witten:1992fb} it is known
that there are topological A-branes which are Lagrangian
submanifolds. This was part of the boundary conditions imposed to
the topological A-model formulated on a Riemann surface with
boundaries. The quantum version of the open topological model was
obtained adding a Wilson loop to the path integral of the closed
topological model. Requiring the Wilson loop to be BRST invariant,
it was found that Lagrangian submanifolds can carry only a flat
vector bundle.

The possibility of a non-flat connection for a D-brane was considered in
\cite{Bershadsky:1995qy, Ooguri:1996ck}. The algebraic
condition that the curvature of a line bundle over the worldvolume of 
the brane has to satisfy in order that the
D-brane is a topological A-brane has been worked out in
\cite{Kapustin:2001ij}. These topological branes are
non-Lagrangian branes of higher dimension than Lagrangian ones and
they have been called coisotropic branes using the language of
symplectic geometry. %If $n$ is the complex dimension of the target
%space manifold, coisotropic A-branes are of real dimensions
%$n+2k$. For $k=0$ they are just Lagrangians branes but for $k \neq
%0$ one has higher dimensional branes.
One of the main motivation to consider such branes was in the
context of homological mirror symmetry. It was argued that
Kontsevich's conjecture \cite{kontsevich} can be true only if the
Fukaya category is enlarged with coisotropic A-branes.

When the target space manifold has a generalized complex geometry
\cite{Hitchin:2002, Gualtieri:2004}, B and A-twisted topological
models have been defined in \cite{Kapustin:2003sg,
Kapustin:2004gv}. The paper \cite{Kapustin:2003sg} also treats the
case of topological branes showing that their geometry is
naturally described in terms of generalized complex submanifolds.
In particular, coisotropic A-branes are generalized complex
submanifolds of a symplectic manifold \cite{Gualtieri:2004}.

In a recent paper \cite{Chiantese:2004pe} mirror symmetry in the
Strominger-Yau-Zaslow (SYZ) picture of T-duality
\cite{Strominger:1996it} is studied in the context of topological
models with generalized complex geometry. Under the explicitly
constructed mirror map topological A and B-branes are mirror
pairs. Furthermore, it was shown that the field strength on the
world volume of the brane is mapped to a bi-vector in the normal
direction of the mirror brane. This leads to a definition of a
generalized tangent bundle when there is a bi-vector in the
normal direction of a generalized submanifold.

In this paper we use the above insight from mirror symmetry to
analyze generalized topological A-branes which are equipped with a
bi-vector in the normal direction of the brane. In this case the
definition of a generalized complex submanifold has to be refined
with respect to the case that there is a two-form on the
submanifold \cite{Gualtieri:2004}. Recall that the two-form
on the submanifold is necessary to have covariance under $B$-field
transformations. This is because a $B$-field transformation acts
as a shear transformation on the cotangent bundle $T^*$. Since a
$\beta$-field transformation acts as a shear transformation on the
tangent bundle $T$, the bi-vector turns out to give covariance
under $\beta$-field transformations. As a natural consequence one
finds generalized complex submanifolds which are isotropic in a
symplectic manifold. The target space of the generalized
topological A-model with complex structures for left and right
movers identified is in particular a symplectic manifold and
therefore one finds isotropic A-branes. Denoting with $n$ the
complex dimension of the target space $N$, this topological branes
are of real dimension $n-2k$, which has to be compared with the
real dimension $n+2k$ of coisotropic A-branes. Isotropic branes
are non-trivial cycles if the Betti numbers $b_{n-2k}(N)$ are non
zero. By Poincar\'e duality $b_{n-2k}(N) = b_{n+2k}(N)$ and
therefore isotropic and coisotropic branes appear in pairs but
with different differential structures on them.

The stability condition for isotropic A-branes is also worked
out following the worldsheet approach applied already to the case
of coisotropic A-branes \cite{Kapustin:2003se}. This is an
important issue because it leads to supersymmetric cycles and to
the existence of BPS states.

The paper is organized as follows. In section (\ref{sec:D2}) we
take a D2 topological B-brane wrapping a two dimensional torus
fiber $T^2$ in the target space $T^6$, which is a trivial $T^3$
torus fibration. We consider this case to construct explicitly the
generalized tangent bundle on the mirror side. In section
(\ref{sec:isotr}) we use the definition of a generalized complex
submanifold equipped with a bi-vector in the normal direction to
show the existence of isotropic A-branes. In section
(\ref{sec:stab}) we work out the condition that a certain form of
the target space has to satisfy in order that isotropic A-branes
are stable. In the last section of the paper we give the
conclusion and outlook.

%%%%%%%%%%%%%%%%%%%%%%%%%%%%%%%%%%%%%%%%%%%%%%%%%%%%%%

\section{D2-brane on the fiber and the mirror brane}
\label{sec:D2}
%%%%%%%%%%%%%%%%%%%%%%%%%%%%%%%%%%%%%%%%%%%%%%%%%%%%%%

Topological A and B-branes are submanifolds $M$ of the target
space $N$ of the twisted $\mathcal N =(2,2)$ supersymmetric non
linear sigma model in two dimensions. The submanifold $M$ is a
topological A or B-brane if the $U(1)$ R-currents
$j_{\pm}=\omega_{\pm}(\psi_{\pm}$, $\psi_{\pm})$ match on
the boundary of the Riemann surface as explained below 
\cite{Kapustin:2003sg}. Note
that we allow for the most general target space geometry, which is
a bi-Hermitian geometry described in terms of two different
complex structures $I_+$ and $I_-$ for right and left
movers.\footnote{Bi-Hermitian means that the Riemannian metric $G
\in \odot^2 T^*_N$ is Hermitian with respect to $I_+$ and $I_-$.
There are also two symplectic structures $\omega_{\pm} = G
I_{\pm}$.} This geometry was first discovered in
\cite{Gates:1984nk} and later included
%seen to fit
in the framework of Hitchin's generalized complex geometry
\cite{Hitchin:2002} by Gualtieri \cite{Gualtieri:2004}. The
conditions $j_+ + j_- = 0$ and $j_+ - j_- = 0$ lead to topological
A and B-branes respectively and preserve $\mathcal N = 2$
worldsheet supersymmetry on the boundary. One has also to require
boundary conditions for the fermions arising from $\mathcal N =1$
worldsheet supersymmetry, which are described in terms of a
gluing matrix $R$ and take the form $\psi_- = R \psi_+$.
The full set of boundary conditions
preserving $\mathcal N = 1$ worldsheet supersymmetry have been
studied in \cite{Albertsson:2001dv, Albertsson:2002qc}. If there
exists a two form $F \in \Lambda^2 T^*_M$ and $I_+ = I_-$, the
matrix $R$ satisfies the equation $\pi^t(G-F)\pi = \pi^t(G+F)\pi
R$, where $\pi$ is the projector on the Neumann directions of the
brane.
Using the form of the projector $\pi$ in \cite{Albertsson:2002qc},
one finds (see \emph{e.g.} \cite{Kapustin:2001ij})
$R^{-1} = (-id_{N_M}) \oplus (g-F)^{-1}(g+F)$ on $T_N|_M
= N_M \oplus T_M$, where $N_M$ is the normal bundle to $M$ 
and $g$ is the restriction of the full metric $G$ to $T_M$.

When $I_+ \neq I_-$ the geometry of topological branes is naturally
described using the language of generalized complex geometry
\cite{Kapustin:2003sg}. The brane geometry was then described
in terms of a gluing matrix $\mathcal R: T_N \oplus T^*_N \to
T_N \oplus T^*_N$ in \cite{Zabzine:2004dp}.
The gluing condition becomes $\Psi = \mathcal R
\Psi$ with
\begin{equation}
\label{eq:gluing}
\R=\begin{pmatrix}
1 & \\ F & 1
\end{pmatrix}
\begin{pmatrix}
r &  \\ & -r^t
\end{pmatrix}
\begin{pmatrix}
1 & \\ -F & 1
\end{pmatrix}
=
\begin{pmatrix}
r & \\ F\, r\,+\, r^t\, F & -r^t
\end{pmatrix} \, .
\end{equation}
In the last equation $r = \pi - Q$, where $Q$ is the projector 
on the Dirichlet directions of the brane and
\begin{equation} \label{bigpsidef}
\Psi := \left( \begin{array}{c}
\psi \\
\rho \end{array}\right) \, , \qquad
\psi := \frac{1}{2}(\psi_+ + \psi_-) \in T_N , \qquad
\rho := \frac{1}{2}G(\psi_+ - \psi_-) \in T^*_N \, .
\end{equation}
Using the projector ${\mathcal R}_+ = (1+\mathcal R)/2$ and
defining
\begin{equation}
\label{eq:tauplus}
\tau_+ : = \big\{X + \xi \in (T_N \oplus
T^*_N)|_M : {\mathcal R}_+ (X+\xi) = X+\xi \big\} \,,
\end{equation}
one finds
\begin{equation}
\label{eq:gentg-marco} \tau_+ \equiv \tau^F_M = \big\{X + \xi \in
T_M \oplus T^*_N|_M : \xi|_M = X\haken F \big\} \,.
\end{equation}
$\tau^F_M$ is the generalized tangent bundle of Gualtieri
\cite{Gualtieri:2004}. By the conditions $j_+ \pm j_- = 0$ written
in terms of the $T_N \oplus T^*_N$ bundle, topological A (B)
branes are described in terms of the generalized K\"ahler
structure $\mathcal J_2$ ($\mathcal J_1$) defined in
\cite{Gualtieri:2004}. Since the matrix $\mathcal R$ commutes with
$\mathcal J_2$ ($\mathcal J_1$) for topological A (B) branes,
topological branes are generalized complex submanifolds $(M,F)
\subset (N, \mathcal J_{2/1})$ according to the definition given
in \cite{Gualtieri:2004}.

The rest of the section is dedicated to show that by applying the SYZ
picture of mirror symmetry to topological branes, one obtains a
new kind of generalized tangent bundle. This definition was
already given in \cite{Chiantese:2004pe}. In that paper, the
target space is a six dimensional torus considered as a trivial
$T^3$ torus fibration. Thus, $N = \mathcal B \oplus \mathcal F :=
T^3 \oplus T^3$, where $\mathcal B$ and $\mathcal F$ are the base
and fiber spaces. The case of a D3 topological A-brane wrapping
$\mathcal F$ was considered.\footnote{We use a notation where the
worldvolume of a Dp-brane is p-dimensional.} Then, under the
explicitly constructed mirror map the D3-brane is mapped to a D0
topological B-brane and the field strength $F$ on the worldvolume
of the D3-brane is mapped to a bi-vector $\tilde{\beta}=F^{-1}$.
This bi-vector lives on the fiber, which is now the normal
direction of the D0-brane.

Here, we want to consider the case of a D2 topological B-brane to
make the appearance of the new generalized tangent bundle explicit. 
Consider the torus fiber as $\mathcal F = \mathcal F_1
\oplus \mathcal F_2 := T^2 \oplus T$. Then, wrap the two
dimensional torus $T^2$ with a D2 topological B-brane having a
field strength on the worldvolume, \emph{i.e.} $F \in \Lambda^2
T^*_{\mathcal F_1}$. The gluing matrix $\mathcal R$ in
(\ref{eq:gluing}) becomes
\begin{equation}
\mathcal R =
\begin{pmatrix}
-1 & 0 & 0 & 0 & 0 & 0 \\
0 & 1 & 0 & 0 & 0 & 0 \\
0 & 0 & -1 & 0 & 0 & 0 \\
0 & 0 & 0 & 1 & 0 & 0 \\
0 & 2 F & 0 & 0 & -1 & 0 \\
0 & 0 & 0 & 0 & 0 & 1
\end{pmatrix} \,,
\end{equation}
which commutes with $\mathcal J_1$. The mirror map
\begin{equation}
\mathcal M : T_{\mathcal B} \oplus T_{\mathcal F_1} \oplus T_{\mathcal F_2}
\oplus T^*_{\mathcal B} \oplus T^*_{\mathcal F_1} \oplus T^*_{\mathcal F_2}
\to T_{\mathcal B} \oplus T^*_{\mathcal F_1} \oplus T^*_{\mathcal F_2}
\oplus T^*_{\mathcal B} \oplus T_{\mathcal F_1} \oplus T_{\mathcal F_2}
\end{equation}
is realized explicitly as
\begin{equation}
\mathcal M =
\begin{pmatrix}
1 & 0 & 0 & 0 & 0 & 0 \\
0 & 0 & 0 & 0 & 1 & 0 \\
0 & 0 & 0 & 0 & 0 & 1 \\
0 & 0 & 0 & 1 & 0 & 0 \\
0 & 1 & 0 & 0 & 0 & 0 \\
0 & 0 & 1 & 0 & 0 & 0
\end{pmatrix} %\,,
\end{equation}
under which the gluing matrix gets mapped to
\begin{equation}
\hat{\mathcal R} = \mathcal M \mathcal R \mathcal M^{-1} =
\begin{pmatrix}
-1 & 0 & 0 & 0 & 0 & 0 \\
0 & -1 & 0 & 0 & 2 \tilde{\beta} & 0 \\
0 & 0 & 1 & 0 & 0 & 0 \\
0 & 0 & 0 & 1 & 0 & 0 \\
0 & 0 & 0 & 0 & 1 & 0 \\
0 & 0 & 0 & 0 & 0 & -1
\end{pmatrix}
\end{equation}
with $\tilde{\beta}=F^{-1} \in \Lambda^2 T_{\mathcal F_1}$. The
above mirror map also exchanges $\mathcal J_1$ with $\mathcal J_2$
\cite{Chiantese:2004pe} and therefore $\hat{\mathcal R}$ commutes
with $\mathcal J_2$. We see that Neumann and Dirichlet boundary
conditions are interchanged so that the D2 topological B-brane is
mapped to a D1 topological A-brane wrapping $\mathcal F_2$. The
field strength $F$ is mapped to a bi-vector $\tilde{\beta}$ living
on $\mathcal F_1$, which is now the normal direction of the
D1-brane. Computing the projector $\hat{\mathcal R}_+ := (1+
\hat{\mathcal R})/2$, one obtains
\begin{equation}
\hat{\mathcal R}_+
\begin{pmatrix}
x_B \\
x_1 \\
x_2 \\
\xi_B \\
\xi_1 \\
\xi_2
\end{pmatrix} =
\begin{pmatrix}
0 \\
\tilde{\beta}(\xi_1) \\
x_2 \\
\xi_B \\
\xi_1 \\
0
\end{pmatrix} \,,
\end{equation}
where $x_1 \in T_{\mathcal  F_1}$ and $x_2 \in T_{\mathcal F_2}$
(the $\xi$'s are the dual coordinates).
Thus, one sees that defining
\begin{equation}
\rho^{\tilde{\beta}}_{\mathcal F_2} := \big\{X+\xi \in (T_N
\oplus T^*_N)|_{\mathcal F_2} :
\hat{\mathcal R}_+ (X+\xi) =
X+\xi \big\} \,,
\end{equation}
we have
\begin{equation}
\rho^{\tilde{\beta}}_{\mathcal F_2} = \big\{ X+ \xi \in
T_{\mathcal F}|_{\mathcal F_2}
\oplus (T^*_{\mathcal B} \oplus T^*_{\mathcal F_1})|_{\mathcal F_2}
: x_1 = \tilde{\beta} (\xi_1)\big\} \,,
\end{equation}
which is stable under $\mathcal J_2$.

The study of the torus indicates that the D2 topological B-brane
$(M,F) \subset (N, \mathcal J_1)$
-- $\tau^F_M$ is stable under $\mathcal J_1$ -- is mapped under the
mirror map to a D1 topological A-brane $(\hat M, \tilde{\beta}) \subset
(\hat N, \mathcal J_2)$ with $\tilde{\beta}= F^{-1} \in \Lambda^2
N_{\hat M}$ and\footnote{$N_M$ is the normal bundle of $M$ in $N$
defined as the quotient $T_N|_M/T_M$. Note that the conormal
bundle $N^*_M$ is the annihilator $\Ann T_M$.}
\begin{equation}
\label{eqn:gent}
\rho^{\tilde{\beta}}_{\hat M} := \big\{ X+ \xi \in T_{\hat N}|_{\hat M}
\oplus N^*_{\hat M} \,:\, X|_{N^*_{\hat M}} = \tilde{\beta}(\xi)\big\}
\end{equation}
stable under $\mathcal J_2$. This lead us to the following
definition. The generalized submanifold $(M, \tilde{\beta})
\subset (N, \mathcal J)$, with $\tilde{\beta} \in \Lambda^2 N_M $,
is a generalized complex submanifold if $\rho^{\tilde{\beta}}_M$
is stable under $\mathcal J$. In the next section we will see that
with this notion of a generalized complex submanifold, in the symplectic
case one finds isotropic A-branes.

%%%%%%%%%%%%%%%%%%%%%%%%%%%%%%%%%%%%%%%%%%%%%%%%%%%%%%%%%%%%%%%%%%%%
\section{Isotropic A-branes}
\label{sec:isotr}
%%%%%%%%%%%%%%%%%%%%%%%%%%%%%%%%%%%%%%%%%%%%%%%%%%%%%%%%%%%%%%%%%%%%%

In this section we want to use generalized complex geometry
to show that there are generalized complex submanifolds of
a symplectic manifold which are isotropic. To this end we need
the definition of a generalized complex submanifold given in
the last section, which we repeat here for clarity.
\begin{definition}
The generalized tangent bundle of a generalized submanifold
$(M,\tilde{\beta})$, with $\tilde{\beta} \in \Lambda^2 N_M $,
is defined by
\begin{displaymath}
\rho^{\tilde{\beta}}_M := \big\{ X+ \xi \in T_N|_M
\oplus N^*_M \,:\, X|_{N^*_M} = \tilde{\beta}(\xi)\big\} \,.
\end{displaymath}
\end{definition}
\begin{definition}
The generalized submanifold  $(M, \tilde{\beta}) \subset (N, \mathcal J)$,
with $\tilde{\beta} \in \Lambda^2 N_M $, is a generalized complex
submanifold if $\rho^{\tilde{\beta}}_M$ is stable under $\mathcal J$.
\end{definition}%
When $M$ is equipped with a 2-form $F \in \Lambda^2 T^*_M$, the
corresponding definitions are given in
\cite{Gualtieri:2004}. In this case the generalized tangent bundle
is given by (\ref{eq:gentg-marco}), which transforms naturally
under $B$-field transformations. In this paper we consider $M$
equipped with a bi-vector $\tilde{\beta} \in \Lambda^2 N_M $ and
therefore the generalized tangent bundle is
$\rho^{\tilde{\beta}}_M$, which transforms naturally under
$\beta$-field transformations.

{}From now on the conormal bundle $N^*_M$ is denoted with the
annihilator $\Ann T_M$. For the concepts of symplectic geometry
used in this section we refer the reader to \cite{daSilva}. Let
$N$ be a symplectic manifold endowed with the symplectic structure
$\omega$ and $(M, \tilde{\beta}) \subset (N, \mathcal J_{\omega})$
its generalized complex submanifold with respect to the
generalized complex structure
\begin{equation}
\mathcal J_{\omega} =
\begin{pmatrix}
 & - \omega^{-1} \\
\omega &
\end{pmatrix} \,.
\end{equation}
Note that for generalized topological A-branes we have to consider
$\mathcal J_2$, which reduces to the symplectic case $\mathcal
J_{\omega}$ for topological A-branes (in Witten's sense of
\cite{Witten:1992fb} where $I_+ = I_-$). Choosing $\beta|_{\Ann
T_M} = \tilde{\beta}$, the bundle $\rho^{\tilde{\beta}}_M$ is
stable under $\mathcal J$ iff
\begin{equation}
\label{eq:beta-rho} e^{-\beta} \rho^{\tilde{\beta}}_M = \rho^0_M =
T_M \oplus \Ann T_M
\end{equation}
is stable under $e^{-\beta} \mathcal J e^{\beta}$.
Here,
\begin{equation}
e^{\beta} =
\begin{pmatrix}
1 & \beta \\
  &   1
\end{pmatrix}
\end{equation}
is the $\beta$-transform of Gualtieri \cite{Gualtieri:2004}.
Requiring the stability of (\ref{eq:beta-rho}) under
\begin{equation}
e^{-\beta} \mathcal J_{\omega} e^{\beta} =
\begin{pmatrix}
- \beta \omega & - \omega^{-1} - \beta \omega \beta \\
\omega & \omega \beta
\end{pmatrix} \,,
\end{equation}
one obtains the following conditions:
\begin{itemize}
\item[1)] \hspace{1mm} $\omega : T_M \to \mathrm{Ann} T_M$,
          \emph{i.e.} M is an isotropic manifold,
\item[2)] \hspace{1mm} $\omega \beta : \mathrm{Ann} T_M \to \mathrm{Ann} T_M$,
\item[3)] \hspace{1mm} $\beta \omega : T_M \to T_M$,
\item[4)] \hspace{1mm} $\omega^{-1} + \beta \omega \beta :
           \mathrm{Ann} T_M \to T_M$.
\end{itemize}
Let us define the symplectic complement of $T_M$ in $T_N|_M$:
\begin{equation}
T^{\omega}_M := \big\{ v \in T_N|_M : \omega(v,w)=0 \;\:
\forall w \in T_M \big\} \,.
\end{equation}
The annihilators $\Ann T_M$ and $\Ann T^{\omega}_M$ are defined as
follows:
{\setlength\arraycolsep{2pt}
\begin{eqnarray}
\label{def:Ann} \Ann T_M & := & \big\{ \eta \in T^*_N|_M :
\eta(v)=0 \;\: \forall
v \in T_M \big\} \,, \nonumber \\
\Ann T^{\omega}_M & := & \big\{ \eta \in T^*_N|_M : \eta(v)=0 \;\:
\forall v \in T^{\omega}_M \big\} \,.
\end{eqnarray}}
Under the isomorphism
\begin{equation}
\label{eq:iso}
\begin{matrix}
(T_N)_x & \to & (T^*_N)_x \\
v & \mapsto & v \haken \omega(x)
\end{matrix} \,,
\end{equation}
the bundle $T^{\omega}_M$ is identified with the annihilator $\Ann
T_M$. This means that $\forall v_{\eta} \in (T^{\omega}_M)_x$ with
$x \in M$, there exists one $\eta \in \Ann (T_M)_x$ such that $
\eta = v_{\eta} \haken \omega(x)$. One also finds that:
\begin{itemize}
\item If $M$ is isotropic in $N$, $T_M \subset T^{\omega}_M$ which
implies that $\Ann T^{\omega}_M \subset \Ann T_M$. \item If $M$ is
coisotropic in $N$, $T^{\omega}_M \subset T_M$ which implies that
$\Ann T_M \subset  \Ann T^{\omega}_M$. \item If $M$ is Lagrangian
in $N$, $T_M = T^{\omega}_M$ which implies that $\Ann T_M = \Ann
T^{\omega}_M$.
\end{itemize}

We want to show that the first and second conditions above lead to
$\tilde{\beta}|_{\Ann T^{\omega}_M} =0 $. The second one tells us
that $\omega \tilde{\beta}(\xi) = \eta$ with $\eta \in
\mathrm{Ann}T_M$, $\forall \xi \in \mathrm{Ann}T_M$. The
nondegeneracy of the symplectic structure in $N$ yields
$\tilde{\beta}(\xi,\sigma) = \omega^{-1}(\eta,\sigma)$ $\forall
\xi,\sigma \in \mathrm{Ann}T_M$ and $\eta \in \Ann T_M$. By means
of the isomorphism (\ref{eq:iso}), the last equation can be
rewritten as $\tilde{\beta}(\xi,\sigma) = \sigma(v_{\eta})$
$\forall \xi,\sigma \in \mathrm{Ann}T_M$ and $v_{\eta} \in
T^{\omega}_M$ . Using the isotropic property $\Ann T^{\omega}_M
\subset \Ann T_M$ and the definition (\ref{def:Ann}) of $\Ann
T^{\omega}_M$, one obtains $\tilde{\beta}(\xi,\sigma) = 0$
$\forall \xi \in \Ann T_M$ and $\forall \sigma \in \Ann
T^{\omega}_M$. We can use again the property $\Ann T^{\omega}_M
\subset \Ann T_M$ and get $\tilde{\beta}(\xi,\sigma) = 0$ $\forall
\xi, \sigma \in \Ann T^{\omega}_M$, which just means that
$\tilde{\beta}|_{\Ann T^{\omega}_M} =0$. Note that this condition
implies that $\tilde{\beta} = 0$ if $M$ is a Lagrangian submanifold.

Defining $E := T^{\omega}_M/T_M$, the consequence of
$\tilde{\beta}|_{\Ann T^{\omega}_M} =0$ is that
$\omega|_E \tilde{\beta}$ becomes an almost complex
structure on $E^*$. In fact, given the natural isomorphism $\Ann T_M
\simeq \Ann T^{\omega}_M \oplus E^*$, $\tilde{\beta} \in
\Lambda^2 N_M$ descends to a bi-vector $\tilde{\beta} \in \Lambda^2 E$.
Furthermore, using the fourth condition above, one finds
$\omega|_E \tilde{\beta} : E^* \to E^*$, \emph{i.e.} $\omega|_E
\tilde{\beta}$ is an almost complex structure on $E^*$.

One can show that $\tilde{\beta} \pm i \omega^{-1}|_E$ are
respectively nondegenerate $(2,0)$ and $(0,2)$ vectors with
respect to the almost complex structure on $E^*$. Therefore, $E$
is of even complex dimension, \emph{i.e.} $\mathrm{dim}|_{\mathbb
R}(T^{\omega}_M/T_M)_x = 2 \mathrm{dim}|_{\mathbb
C}(T^{\omega}_M/T_M)_x = 4k$. By the dimension theorem for the
symplectic complement, we have $\mathrm{dim}|_{\mathbb R} (T_N)_x
= \mathrm{dim}|_{\mathbb R}(T_M)_x + \mathrm{dim}|_{\mathbb
R}(T^{\omega}_M)_x$ and in addition $\mathrm{dim}|_{\mathbb R}
(T^{\omega}_M)_x = \mathrm{dim}|_{\mathbb R}(T^{\omega}_M/T_M)_x +
\mathrm{dim}|_{\mathbb R}(T_M)_x$. Therefore,
$\mathrm{dim}|_{\mathbb R} M = n - 2k$ with $n =
(\mathrm{dim}|_{\mathbb R} N)/2$. Note that $k=0,1, \ldots
n/2$ for $n$ even and $k=0,1, \ldots (n-1)/2$ for $n$ odd.
For $k=0$, $\mathrm{dim}|_{\mathbb R} M = n$ and $M$ is a 
Lagrangian submanifold.

The results above show the existence of generalized complex
submanifolds $M$ of real dimensions $n-2k$ which are isotropic in
the symplectic manifold $N$ of real dimension $2n$.
We recall that the topological B-model is well defined at the
quantum level if $N$ is a Calabi-Yau manifold but for the
topological A-model $N$ can also be a K\"ahler manifold. The
conclusions above imply that there could be isotropic A-branes in
addition to the Lagrangian A-branes of Witten \cite{Witten:1992fb}
and coisotropic A-branes of Kapustin and Orlov
\cite{Kapustin:2001ij}. Since we want to see under which
conditions there are isotropic A-branes besides Lagrangian ones,
we shall take $N$ to be a K\"ahler or Calabi-Yau manifold\footnote{For
a Calabi-Yau $n$-fold we denote, perhaps with a slight abuse of teminology,
a K\"ahler manifold with holonomy group smaller or equal to 
$\mathrm{SU}(n)$.} and $k \neq 0$. 
If all Betti numbers $b_{n-2k}(N)$ are equal to zero,
there are no isotropic A-branes because they are homologically
trivial. However, one has to expect their appearance whenever
$b_{n-2k}(N) \neq 0$. If $M'$ is a coisotropic A-brane,
$\mathrm{dim}|_{\mathbb R} M' = n + 2k$. By Poincar\'e duality,
$b_{n-2k}(N) = b_{n+2k}(N)$ and the same considerations about
homological triviality for coisotropic branes
\cite{Kapustin:2001ij} apply to isotropic branes. For instance, we
know that $b_1(N) = 0$ if $N$ is a compact Riemannian manifold
with holonomy group $\mathrm{SU}(n)$ \cite{Joyce}. Therefore, there are no
one dimensional isotropic A-branes if $N$ is a compact odd
dimensional Calabi-Yau manifold with holonomy $\mathrm{SU}(n)$. This means
that the only compact Calabi-Yau 3-folds $N$ which allow isotropic
A-branes are $N=T^6$ and $N=T^2 \times K3$, which have holonomy
group smaller than  $\mathrm{SU}(3)$. There are also no isotropic A-branes for
complete intersection Calabi-Yau manifolds of odd dimensions in a
projective space because all odd Betti numbers, with the exception
of $b_n(N)$ corresponding to Lagrangian submanifolds, are zero.
The same consideration does not apply when the dimension is even
because the even Betti numbers are different from zero and
therefore there are isotropic A-branes.

%%%%%%%%%%%%%%%%%%%%%%%%%%%%%%%%%%%%%%%%%%%%%%%%%%%%%%%%%%%%%%%%%%%%
\section{Stability condition for isotropic A-branes}
\label{sec:stab}
%%%%%%%%%%%%%%%%%%%%%%%%%%%%%%%%%%%%%%%%%%%%%%%%%%%%%%%%%%%%%%%%%%%%%

The aim of this section is to compute the stability condition for
isotropic A-branes following the worldsheet approach used in
\cite{Kapustin:2003se} to find stability for coistropic A-branes.
In section (\ref{sec:D2}), when $M$ is equipped with a 2-form $F
\in \Lambda^2 T^*_M$, the gluing matrix $\mathcal R: T_N \oplus
T^*_N \to T_N \oplus T^*_N$ was constructed from the boundary
conditions written in terms of the gluing matrix $R: T_N \to T_N$
satisfying the equation $\pi^t(G-F)\pi = \pi^t(G+F)\pi R$. Here,
the manifold $M$ is equipped with a bi-vector $\tilde{\beta} \in
\Lambda^2 N_M $ and we follow the opposite approach. We first
construct the gluing matrix $\mathcal R$ which leads to a
generalized tangent bundle $\tau_+$ defined in (\ref{eq:tauplus})
equal to the generalized tangent bundle $\rho^{\tilde{\beta}}_M$
defined in the previous section. Then, we find the condition the
gluing matrix $R$ has to satisfy by projecting on the $T_N$
bundle.

In the paper \cite{Albertsson:2001dv} the $\mathcal N =1$ boundary
conditions were studied when $F=0$. In a following paper
\cite{Albertsson:2002qc} the case with $F \neq 0$ was considered.
In the latter paper the Dirichlet boundary conditions are unmodified
with respect to the case $F=0$. It was required that $Q R = R Q =
-Q$, where Q is the projector on the Dirichlet directions, \emph{i.e.}
\begin{equation}
Q =
\begin{pmatrix}
  &                \\
  & \delta^i_{\;j}
\end{pmatrix} \,, \qquad \quad
R =
\begin{pmatrix}
R^m_{\;\;\; n} &             \\
          & - \delta^i_{\;j}
\end{pmatrix} \,.
\end{equation}
The indices $i,j$ and $m,n$ label the Dirichlet and Neumann directions
respectively. Then, the projector on the Neumann directions is
defined by $\pi = 1 - Q$. In our case with $\tilde{\beta} \neq 0$,
the boundary conditions on the Neumann directions are unmodified
and we require $\pi R = R \pi = \pi$, \emph{i.e.}
\begin{equation}
\pi =
\begin{pmatrix}
\delta^m_{\;\;\; n} & \\
               &
\end{pmatrix} \,, \qquad \quad
R =
\begin{pmatrix}
\delta^m_{\;\;\; n} &           \\
               & R^i_{\;j}
\end{pmatrix} \,.
\end{equation}
Then, the projector on the Dirichlet directions is defined by
$Q = 1 - \pi$.

One can check that $\tau_+ = \rho^{\tilde{\beta}}_M$ is given by
\begin{equation}
\label{eq:Rbeta}
\R=\begin{pmatrix}
1 & \tilde{\beta} \\
  &      1
\end{pmatrix}
\begin{pmatrix}
r &      \\
  & -r^t
\end{pmatrix}
\begin{pmatrix}
1 & -\tilde{\beta} \\
  &       1
\end{pmatrix}
=
\begin{pmatrix}
r &  - r\, \tilde{\beta}\,  - \, \tilde{\beta} \, r^t \\
  &                   -r^t
\end{pmatrix} \, ,
\end{equation}
where $r = \pi - Q$. The last expression can be compared with the
one in (\ref{eq:gluing}) to see that now $\mathcal R$ is an upper
triangular matrix while before it was a lower triangular matrix.
It was already suggested in \cite{Zabzine:2004dp} to relax the
condition that $\mathcal R$ is a lower triangular matrix, but no
explicit form was given. Here, the form (\ref{eq:Rbeta}) for the
matrix $\mathcal R$ is a natural consequence of the definition of
the generalized tangent bundle $\rho^{\tilde{\beta}}_M$. Using the
gluing conditions $\Psi = \mathcal R \Psi$ and $\psi_- = R
\psi_+$, one obtains
\begin{equation}
Q(G^{-1} - \tilde{\beta})Q^t + Q(G^{-1} + \tilde{\beta})Q^t R^t =
0 \, .
\end{equation}
The explicit expression for the projector $Q$ and the property
$\tilde{\beta}|_{\Ann T^{\omega}_M} =0$ can be used to get
\begin{equation}
\label{eq:Rinv}
(R^{-1})^t = id_{T^*_M} \oplus (- id_{\Ann
T^{\omega}_M}) \oplus \big\{- (g^{-1} - \tilde{\beta})^{-1} (g^{-1}
+ \tilde{\beta})\big\}
\end{equation}
acting on $T^*_N|_M \simeq T^*_M \oplus \Ann T^{\omega}_M \oplus
E^*$. The metric $g$ is now the restriction of the full metric $G$ 
to $N_M$. Here, we are concerned with ordinary topological
A-branes for which $I_+ = I_- = I$. The boundary condition $j_+ +
j_- =0$, where $j_{\pm} = \omega(\psi_{\pm}, \psi_{\pm})$, gives
$R^t \omega R = - \omega$. Using $\omega = G I$ and $R^t G R = G$,
one finds that the matrix $R$ anticommutes with the complex
structure $I$ of the target space manifold $N$. This means that
$R$ is of the form
\begin{equation}
R =
\begin{pmatrix}
    & R_1 \\
R_2 &
\end{pmatrix}
\end{equation}
with $R_1 : T^{(0,1)}_N \to T^{(1,0)}_N$ and $R_2 : T^{(1,0)}_N
\to T^{(0,1)}_N$.

The stability condition for topological branes can be derived by
requiring that the spectral flow operators
\begin{equation}
S_{\pm} = \Omega( \psi_{\pm}, \ldots , \psi_{\pm}) = \frac{1}{n!}
\Omega_{i_1 \cdots i_n} \psi^{i_1}_{\pm} \cdots \psi^{i_n}_{\pm}
\end{equation}
match properly on the boundary of the Riemann surface.\footnote{We
use the conventions of \cite{Witten:1991zz} where the BRST
operator of the topological A-model is $Q_A = Q_+ + \bar Q_-$ and
$\psi^i_+$ and $\psi^{\bar i}_-$ are the scalar fields on the
Riemann surface.} Here, $N$ is a Calabi-Yau manifold and $\Omega$
is the holomorphic $(n,0)$-form. For topological A-branes the
stability condition is
\begin{equation}
S_+ = e^{i\alpha} \bar S_-
\end{equation}
with $\alpha$ a real constant. Using the gluing condition $\psi_+ =
R^{-1}_2 \psi_-$, the last equation yields
\begin{equation}
\label{eq:omega-det}
\Omega_{1 \cdots n} \det (R^{-1}_2) = e^{i\alpha} \bar \Omega_{1
\cdots n} \, .
\end{equation}
The equation (\ref{eq:Rinv}) gives
\begin{equation}
\label{eq:det}
\det (R^{-1}_2) = (-1)^{p/2} (-1)^{2k} \det \big\{(G^{-1} -
\tilde{\beta})^{-1} (G^{-1} + \tilde{\beta})|_{E^{* (1,0)}} \big\}
\, ,
\end{equation}
where we have used that $(R^{-1}_2)^t : T^{*(1,0)}_N \to
T^{*(0,1)}_N$ and that $\mathrm{dim}|_{\mathbb R} (\Ann
T^{\omega}_M)_x = \mathrm{dim}|_{\mathbb R} (T^*_M)_x = p$ with $p =
n - 2k$. Let us define the following matrices
\begin{equation}
\label{eq:mxbeta}
G|_E =
\begin{pmatrix}
    & g \\
g^t &
\end{pmatrix} \, , \qquad \quad
\tilde{\beta} =
\begin{pmatrix}
\tilde{\beta}_{(2,0)} & \tilde{\beta}_{(1,1)} \\
- \tilde{\beta}_{(1,1)}^t & \tilde{\beta}_{(0,2)}
\end{pmatrix}\, ,
\end{equation}
where now $g : E^{(0,1)} \to E^{*(1,0)}$ and $\tilde{\beta} : E^*
\to E$. These matrices are used to rewrite the matrix in equation
(\ref{eq:det}) in terms of $(p,q)$-type components of $\tilde{\beta}$.
To compute the determinant in (\ref{eq:det}) we need the following
identities
{\setlength\arraycolsep{2pt}
\begin{eqnarray}
\label{eq:identbeta}
g \tilde{\beta}_{(0,2)} & = & - (1 - g
\tilde{\beta}^t_{(1,1)}) (g^t \tilde{\beta}_{(2,0)})^{-1}(1 - g^t
\tilde{\beta}_{(1,1)}) \, ,
\nonumber \\
g^t \tilde{\beta}_{(2,0)} & = & - (1 - g^t \tilde{\beta}_{(1,1)})
(g \tilde{\beta}_{(0,2)})^{-1}(1 - g \tilde{\beta}^t_{(1,1)}) \, ,
\end{eqnarray}}
which can be derived by the fact that $\omega|_E \tilde{\beta}$
is an almost complex structure on $E^*$, \emph{i.e.}
$(\omega|_E \tilde{\beta})^2 = -1$. Then, one obtains
\begin{equation}
(G^{-1} - \tilde{\beta})^{-1}
(G^{-1} + \tilde{\beta})|_{E^{* (1,0)}} =
(1 - g^t \tilde{\beta}_{(1,1)})^{-1} g^t \tilde{\beta}_{(2,0)} \, ,
\end{equation}
whose determinant is easy to compute. Multiplying the equation
(\ref{eq:omega-det}) by $\Omega_{1 \cdots n}$ and using the
proportionality relation $|\Omega_{1 \cdots n}|^2 \propto
\sqrt{\det G}$, it follows that
\begin{equation}
\label{eq:Osqured}
(\Omega_{1 \cdots n})^2 \det
\tilde{\beta}_{(2,0)} \propto e^{i \alpha} \sqrt{\det G} \det
\big\{ (g^t)^{-1} - \tilde{\beta}_{(1,1)} \big\} \, .
\end{equation}
The identities (\ref{eq:identbeta}) can be used to get
\begin{equation}
\frac{\det \big(G^{-1} + \tilde{\beta}\big)}{\det G^{-1}} = 2^{2k}
\det g \det \big\{ (g^t)^{-1} - \tilde{\beta}_{(1,1)}\big\}
\end{equation}
by which the equation (\ref{eq:Osqured}) is rewritten as
\begin{equation} \label{eq:flowmatch}
(\Omega_{1 \cdots n})^2 \det \tilde{\beta}_{(2,0)} \propto e^{i
\alpha} \frac{\sqrt{\det G}}{\sqrt{\det G|_E}} \frac{\det \big(
G^{-1} + \tilde{\beta}\big)}{\det G^{-1}} \, .
\end{equation}

A Lagrangian A-brane is stable if it is a special Lagrangian
submanifold $M$ of the Calabi-Yau manifold $N$. This means that
$M$ is a calibrated submanifold with respect to the holomorphic
$(n,0)$-form on $N$ or in other words that $M$ is defined using a
closed form on $N$. Following this idea, we want to find a form on
$N$ by which we can define our isotropic submanifolds to be
stable. Then, one can rewrite this form in terms of the geometric
data found before.

Recall that $T^*_N|_M \simeq \Ann T^{\omega}_M \oplus T^*_M
\oplus E^*$. Choose a complex basis $\{f^1, \ldots , f^{p}\}$ for
$\Ann T^{\omega}_M \oplus T^*_M$ such that $\{\Im \mathrm m f^1,
\ldots , \Im \mathrm m f^{p}\}$ and $\{\Re \mathrm e f^1, \ldots
, \Re \mathrm e f^{p}\}$ span $\Ann T^{\omega}_M$ and $T^*_M$
respectively. The complex basis $\{f^{p+1}, \ldots , f^n\}$ can
be chosen to span $E^*$ so that $\{f^1, \ldots , f^n\}$ is a
complex basis for $T^*_N|_M$. In terms of this basis
\begin{equation}
\Omega|_{T^{\omega}_M} = \Omega_{1 \cdots n} \Re \mathrm e f^1
\wedge \cdots \wedge \Re \mathrm e f^{p} \wedge f^{p+1} \wedge
\cdots \wedge f^n \, .
\end{equation}
The last (n,0)-form can be multiplied by the $2k$-form
$(\tilde{\beta}^{-1})^k$ to provide the volume form
$\mathrm{vol}_{T^{\omega}_M}$ on ${T^{\omega}_M}$. The property
$(\omega|_E \tilde{\beta})^2 = -1$ gives
\begin{equation}
\tilde{\beta}^{-1} = \omega|_E \tilde{\beta} \omega^t|_E =
\begin{pmatrix}
  -g \tilde{\beta}_{(0,2)} g^t  & -g \tilde{\beta}^t_{(1,1)} g \\
  g^t \tilde{\beta}_{(1,1)} g^t & -g^t \tilde{\beta}_{(2,0)} g \\
\end{pmatrix} \, ,
\end{equation}
which can also be obtained computing the inverse of the matrix
$\tilde{\beta}$ in (\ref{eq:mxbeta}) and using the identities
(\ref{eq:identbeta}). We will need the definition of the Pfaffian
of a skew-symmetric $2n \times 2n$ matrix, which we recall here.
Consider a $2n$-dimensional Riemannian manifold and a two-form
$\alpha = A_{ab} e^*_a \wedge e^*_b$ associated to a
skew-symmetric $2n \times 2n$ matrix $A$ with elements $A_{ab}$.
The Pfaffian of $A$ is defined by the equation
\begin{equation}
\frac{1}{n!} \alpha^n = \mathrm{Pf}(A) e^*_1 \wedge \cdots \wedge
e^*_{2n} \, .
\end{equation}
Using this definition, we find
{\setlength\arraycolsep{2pt}
\begin{eqnarray}
\label{eq:volume}
\frac{1}{k!} \Omega|_{T^{\omega}_M} \wedge (\tilde{\beta}^{-1})^k
& = & \Omega_{1 \cdots n}
\mathrm{Pf}\big\{(\tilde{\beta}^{-1})_{(0,2)}\big\} \Re \mathrm e
f^1 \wedge \cdots \wedge \Re \mathrm e f^{p} \wedge f^{p+1}
\wedge \cdots \wedge f^n \nonumber \\
& &  \hphantom{ \Omega_{1 \cdots n}
\mathrm{Pf}\big\{(\tilde{\beta}^{-1})_{(0,2)}\big\}}
                 \wedge \bar f^{p+1} \wedge \cdots \wedge
\bar f^n \nonumber \\[2pt]
& = & \Omega_{1 \cdots n}
\frac{\mathrm{Pf}\big\{(\tilde{\beta}^{-1})_{(0,2)}\big\}}{\sqrt{\det
G|_{T^{\omega}_M}}} \mathrm{vol}_{T^{\omega}_M} \, ,
\end{eqnarray}}
where $(\tilde{\beta}^{-1})_{(0,2)}$ is the $(0,2)$-type component
of $\tilde{\beta}^{-1}$, \emph{i.e.} $(\tilde{\beta}^{-1})_{(0,2)}
= -g^t \tilde{\beta}_{(2,0)} g$.

In the last equation we have used the definition of a volume form
given in \cite{Joyce}. If $(N,G)$ is a Riemannian
manifold, an oriented tangent $m$-plane on $N$ is a
$m$-dimensional vector subspace $V$ of $(T_N)_x$ equipped with an
orientation. The volume form $\mathrm{vol}_V$ on $V$ is 
a m-form on $V$, which is given by the combination of the metric
$G|_V$ on $V$ with the orientation on $V$.
%The metric on $V$ is $G|_V$ and the combination of
%$G|_V$ with the orientation on $V$ gives a m-form on $V$, which is
%the volume form $\mathrm{vol}_V$ on $V$. 
In our case $(T^{\omega}_M)_x$ $\forall x \in M$ is an oriented 
tangent $(4k + p)$-plane on $N$. Thus, the combination of
$G|_{(T^{\omega}_M)_x}$ with the orientation on $(T^{\omega}_M)_x$
gives a $(4k + p)$-form on $(T^{\omega}_M)_x$, which is the
volume form $\mathrm{vol}_{(T^{\omega}_M)_x}$ on
$(T^{\omega}_M)_x$ for all $x \in M$. The form in
(\ref{eq:volume}) is a $(n+2k)$-form on $T^{\omega}_M$, which is a
$(4k+p)$-form because $n= p+2k$.

The properties $\mathrm{Pf} (BAB^t) = \det (B) \mathrm{Pf} (A)$
and $\big(\mathrm{Pf} (A)\big)^2 = \det (A)$, where $B$ is an
arbitrary matrix of rank $2n$, are used to rewrite the equation
(\ref{eq:volume}) as
\begin{equation}
\frac{1}{k!} \Omega|_{T^{\omega}_M} \wedge (\tilde{\beta}^{-1})^k
\propto  \Omega_{1 \cdots n}
\frac{\det g \sqrt{\det \tilde{\beta}_{(2,0)}}}{\sqrt{\det
G|_{T^{\omega}_M}}} \mathrm{vol}_{T^{\omega}_M} \, .
\end{equation}
The condition (\ref{eq:flowmatch}) derived from the matching of
the spectral flow operators on the boundary can be used for the
last equation to give the stability condition in terms of a form
on $N$. Using the identity $\det G|_E \det G = (\det
G|_{T^{\omega}_M})^2$, the stability condition for isotropic
A-branes is
\begin{equation} \label{eq:stab}
\frac{1}{k!} \Omega|_{T^{\omega}_M} \wedge (\tilde{\beta}^{-1})^k
\propto e^{i \frac{\alpha}{2}} \sqrt{\frac{\det \big( G^{-1} +
\tilde{\beta}\big)}{\det G^{-1}}} \mathrm{vol}_{T^{\omega}_M} \, .
\end{equation}

The stability condition just derived is in agreement with the
condition for a Lagrangian A-brane to be stable. It is known that
special Lagrangian submanifolds are stable Lagrangian A-branes.
When an isotropic A-brane is a Lagrangian submanifold,
$T^{\omega}_M = T_M$, $k=0$ and $\tilde{\beta}=0$. Therefore, the
equation (\ref{eq:stab}) gives
\begin{equation}
\Omega|_{T_M} \propto e^{i \frac{\alpha}{2}} \mathrm{vol}_{T_M} \,
,
\end{equation}
which is the condition for a Lagrangian submanifold to be special.

Note that the study of stability for topological A-branes with a
non-trivial field strength on the worldvolume of the brane was
performed in \cite{Marino:1999af} using the supersymmetric
Born-Infeld action. The authors of \cite{Kapustin:2003se} 
found agreement with these results using the worldsheet approach. 
However, the case of the five dimensional coisotropic
A-brane in $T^6$ or $T^2 \times K3$ was not analyzed in
\cite{Marino:1999af}. For the isotropic non-Lagrangian A-branes
studied in this paper the case is rather different because the
brane is equipped with a non-trivial bi-vector living in the
normal direction of the brane. For a Calabi-Yau three-fold which
is $T^6$ or $T^2 \times K3$, in addition to three dimensional
Lagrangian A-branes and five dimensional coisotropic A-branes,
there are one dimensional isotropic A-branes, whose stability
condition is given by (\ref{eq:stab}) with $k=1$. For a Calabi-Yau
four-fold $N$ we want to mention the case of
two dimensional isotropic A-branes, whose stability condition is
still given by (\ref{eq:stab}) with $k=1$.

%%%%%%%%%%%%%%%%%%%%%%%%%%%%%
\section{Conclusion and outlook}
%%%%%%%%%%%%%%%%%%%%%%%%%%%%%

In this paper we have shown the existence of isotropic A-branes
whenever they are homological non-trivial cycles. We have also
given the stability condition for these cycles by the worldsheet
approach. Using Poincar\'e duality, it was argued that coisotropic
and isotropic A-branes are dual cycles but with different
differential structures on them. The coisotropic A-branes are
equipped with a non-trivial curvature for the connection of a line bundle
on the worldvolume of the brane while the isotropic ones are equipped 
with a non-trivial
bi-vector in the normal direction of the brane. In the paper
\cite{Kapustin:2001ij} it was argued that if Kontsevich's
conjecture is true, the Fukaya category should be enlarged with
coisotropic A-branes. It is then natural to propose that isotropic
A-branes give a further enlargement of the Fukaya category.

Kontsevich's conjecture is about the equivalence of two
triangulated categories for mirror manifolds $N$ and $\hat N$:
the derived category of coherent sheaves $D^b(N)$ and the derived 
Fukaya category $DF(\hat N)$. In physical terms  $D^b(N)$ is the category 
of topological B-branes while $DF(N)$ is the category of Lagrangian 
A-branes. For a Lagrangian A-brane to belong to $DF(N)$ one has 
to require an anomaly cancellation
condition. We recall that at the classical level the boundary
conditions for topological A-branes have been defined to preserve
the axial R-symmetry, which could be broken at the quantum level.
It turns out that the condition for a Lagrangian A-brane to be
special corresponds to the vanishing of the Maslov class, which is
responsible for the anomaly. In mathematical terms this means that
the Lagrangian submanifold is gradable in the sense of Kontsevich.
The anomaly free condition for coisotropic A-branes was studied in
\cite{Li:2004qh} based on a proposal from the stability condition
worked out in \cite{Kapustin:2003se}. The stability condition for
coisotropic A-branes dictates a generalization of the Maslov
class, whose zero gives the anomaly cancellation and graded
coisotropic A-branes. It is clear that an analogous study for
isotropic A-branes should be done based on the stability condition
worked out here. This would correspond to graded isotropic
A-branes, which together with the graded coisotropic ones might
give the full enlargement of the Fukaya category from a physics 
perspective.

%%%%%%%%%%%%%%%%%%%%%%%%%%%%%
\section*{Acknowledgements}
%%%%%%%%%%%%%%%%%%%%%%%%%%%%%

I would like to thank the string theory group of the University of
Wisconsin-Madison for hospitality during the preparation of part 
of this work. In particular I am grateful to Albrecht Klemm for 
several interesting discussions. I also would like to thank
Marco Gualtieri for useful discussions and for pointing out 
a relevant paper. Thanks also to Florian Gmeiner and Claus Jeschek 
for a recent collaboration and to Ingo Runkel for reading the manuscript.
This work is supported by DFG Graduiertenkolleg 271/3-02.

% ---- Bibliography ----

%\nocite{*} %this uses *everything* in the .bib file
%\bibliographystyle{utphys} %if you use utphys.bst
%\bibliography{topsigma} %or whatever your .bib file is

%%%%%%%%%%%%%%%%%%%%%% here started the bib ... %%%%%%%%%%%%%%%

%\providecommand{\href}[2]{#2}
%\begingroup\raggedright\begin{thebibliography}{10}

\end{document}